\documentclass[fleqn]{article}
\usepackage{amsmath,amssymb}
\usepackage[dvips]{graphicx}

\usepackage{latexsym}

\begin{document}

\pagestyle{plain} 
\setcounter{page}{1}
\setlength{\textheight}{700pt}
\setlength{\topmargin}{-40pt}
\setlength{\headheight}{0pt}
\setlength{\marginparwidth}{-10pt}
\setlength{\textwidth}{20cm}

\title{$p$-th Clustering coefficients and $q$-$th$ degrees of separation  based on String-Adjacent   Formulation   }
\author{Norihito Toyota   \and Hokkaido Information University, Ebetsu, Nisinopporo 59-2, Japan \and email :toyota@do-johodai.ac.jp }
\date{}
\maketitle

\begin{abstract}
The phenomenon of six degrees of separation is an old but attractive subject.  
The deep understanding  has been uncovered yet, especially how  closed paths included in a network affect  six degrees of separation are an important subject left yet.   
For it, some researches  have been made\cite{Newm21},  \cite{Aoyama}.  
 Recently we have develop a formalism \cite{Toyota3},\cite{Toyota4} to explore the subject  based on the string formalism developed by Aoyama\cite{Aoyama}.  
The formalism can systematically investigate the effect of closed paths, especially generalized clustering coefficient  $C_{(p)}$  introduced in \cite{Toyota4},  on six degrees of separation.     
In this article, we analyze general $q$-th degrees of separation by using 
 the formalism developed by us. 
 So we find that the scale free network with exponent $\gamma=3$ just display six degrees of separation.  
 Furthermore we drive a phenomenological relation between the separation number $q$ and $C_{(p)}$ that has crucial information on circle structures in networks.     
  \end{abstract}
\begin{flushleft}
\textbf{keywords:}
Six Degrees of Separation, String, Clustering Coefficient, Adjacent Matrix, Generalized Clustering Coefficient
\end{flushleft}

\section{Introduction}\label{intro}

\hspace{5mm} In 1967, Milgram has made a great impact on  the world by advocating  
 "six degrees of separation" by  a celebrated paper \cite{Milg} written based on an social experiment. 
 "Six degrees of separation" indicates that people have a narrow circle of acquaintances. 
A series of social experiments made by him and his joint researchers \cite{Milg2},\cite{Milg3} 
made the suggestion, which all people in USA are connected through about 6 intermediate acquaintances, more certain.    

The two breakthroughs have made in the end of last century in network theory that declare
 the start of "complex network theory". 
One is small world networks that have been proposed by  Watts and Strogatz\cite{Watt1},\cite{Watt2}.  
Another is the scale free networks proposed by Barabasi et al.\cite{Albe2}, \cite{Albe3}. 
Many empirical networks exhibit characteristic future of scale free \cite{Albe1},\cite{Newm},\cite{Doro1},\cite{Doro2}. 
Their frameworks provided compelling evidence that the small-world phenomenon is pervasive in a range of networks arising in nature and technology, and a fundamental ingredient in the evolution of the World Wide Web. 
Furthermore Watts and his coworkers continued to explore  six degrees of separation\cite{Watt4},\cite{Watt3}.  
We, however, think that the phenomenon, six degrees of separation, is not understood well in theoretical point of view. 
Especially how does the clustering coefficient proposed in \cite{Watt1} have  an effect on it? 
If the network of human relations has a tree structure without circles,  a person connects new persons  in power  of average degree, when he(she) follows his(her)  acquaintances  step by step on his(her) network of human relations. Then six degrees of separation is not so amazing,  if a person has a few hundred acquaintances. 
A question is that  networks of general human relations include some circles. 
This structures would decrease the number of new persons that connected with him(her) when he(she) follows  his(her) acquaintances  step by step.  
One of indices characterizing  circle structures is the clustering coefficient. 
Thus it will be important to investigate the effect of the clustering coefficient  on six degrees of separation. 
It is, however, difficult  to investigate the influence of circle structures  with general size.    
There are in fact only  a little researches focused on  the effect of circle structures.

We have studied it from theoretical point of view with such motives. 
First we  investigated  it by imposing a homogeneous hypothesis on networks\cite{Toyota1}.  
As a result, we found that the clustering coefficient has not any decisive effects on the propagation 
of information on a network and then information easily spread to a lot of people even in networks with 
 relatively large clustering coefficient under the hypothesis; a person only needs dozens of friends for six degrees of separation.   
Moreover we devoted deep study to the six degrees of separation based on some models proposed 
by  Pool and Kochen \cite{Pool} by using a computer, numerically\cite{Toyota2}. 
 In the article, we  estimated the clustering coefficient along the method  developed by us \cite{Toyota1} 
and improved our analysis of the subject through marrying Pool and Kochen's models to our method introduced in \cite{Toyota1}.   
As a result, it seems to be difficult that six degrees of separation is realized in the models proposed by Pool and Kochen\cite{Pool} on the whole. 

The studies was, however,  made only under rather restricted conditions on networks. 
Newman  studied the influence of circle structures in general networks  on the subject\cite{Newm21}. 
The study is so stimulating  but only triangle structures and quadrilateral structures on networks were considered.     
It seems to be difficult to  generalize  his framework to $p$-polygon 
that are circles with general size $p$.   
Recently Aoyama proposed the string formulation for the  subject\cite{Aoyama}.  
The idea  inspired our study in this article, greatly.  
Although the formalism is available for general networks with any circles,  he unfortunately tacked the subject only at tree approximation  of  networks. 
Since he deals with mainly scale free networks with small clustering coefficient, the approximation is valid up to a certain  point.   
We developed  the string formalism by fusing adjacent matrix formulation so as  one can analyze six degrees of separation even in networks with  general size of circles\cite{Toyota3},\cite{Toyota4}.  

In \cite{Toyota3}, the formalism and the justification of it  are mainly given, and the formalism and 
analyses of two degrees of separation  as  preliminary results were reported in  \cite{Toyota4}.  
Although we also defined the general $p$-Clustering coefficient $C_{(p)}$ in \cite{Toyota4}, we do not  
discuss any relation between six degrees of separation and $C_{(p)}$  yet.  
In this article we pursue  the relations between separation number $q$ and  $C_{(p)}$ as well as general $q$ degrees of separation (where $q\leq 6$) in string formulation.  
After that, we show that some phenomenological relation holds. 
The result  naturally reflects the effect of circle structures in networks on separation.

 The plan of this article is as follows. 
After introduction,  we briefly review the formalism developed in \cite{Toyota3},\cite{Toyota4}
 in the following section 2.  
According the formalism, we introduce generalized $p$-th clustering coefficients as well as the usual global one.   
In the next section 3, 
$q$-th degrees of separation  (where $q\leq 6$)  in scale free networks \cite{Albe2}, \cite{Albe3} 
with various values of the exponents based on Milgram condition proposed by Aoyama\cite{Aoyama}. 
Though the obtained result  is a little different from Aoyama's one,  it is not contradictory to Aoyama's conjecture crucially. 
The justification for our result is provided by estimating the power $A^q$ of an adjacent matrix $A$.    
We  discuss  the relation between the separation number $q$ and  $C_{(p)}$ in the section 4. 
We show a phenomenological relation holds there.  
The last section 5  is devoted to summary.

\section{Review for String Formulation and Adjacent Matrix}
\label{usage}
\subsection{String Formalism}
We review the formalism given in  \cite{Toyota3},\cite{Toyota4} , according to the formulation introduced by Aoyama \cite{Aoyama}. 

 We consider a string-like part of a graph with connected $j$ vertices  and call it  "j-string".
$N$ is the number of vertices in a considering network and  $S_j$ is the number of j-string in the network. 
(Note that  $S_j$ in this article is $N$   times larger than $S_{j-1}^{Aoyama}$ defined by Aoyama\cite{Aoyama}.)     
By definition, $S_1=N$ and $S_2$ is the number of edges  in the network. 
$\bar{S}_j$ is the number of non-degenerate  j-string where a non-degenerate string is defined as strings without any multiple edges and/or any circles in the subgraphs as seen in Fig.1. 
We, however, define that the non-degenerate string contains  strings homeomorphic to a circle.  

We call  strings without any circles as subgraphs and/or whole graphs "open string" and  strings overall homeomorphic to a circle "closed string". 
Thus we consider closed strings and open strings in this article. 

It is so difficult to calculate $S_j$ and $\bar{S}_j$, generally. 
It would be maybe impossible to calculate $S_j$ and $\bar{S}_j$ with $j>7$ explicitly at the present moment.  

 \begin{figure}[t]
%%%%%%%%%%%%%%%%%%%%%%% 
\begin{center}
\includegraphics[scale=0.9,clip]{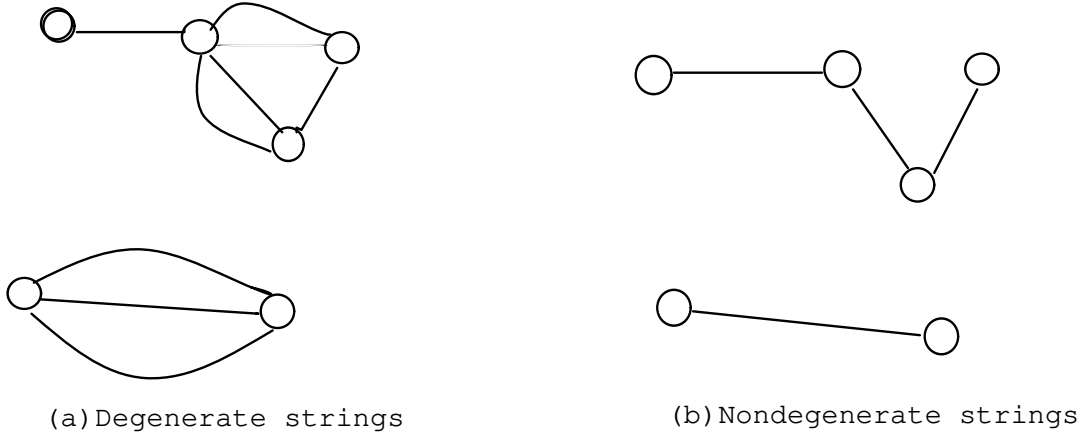} 
\end{center}
\caption{Two types of strings   }
%%%%%%%%%%%%%%%%%%%%
 \end{figure} 
 
\subsection{Generalized Clustering Coefficient}
By using the string formulation, we can defined  the usual clustering coefficient which essentially counts the number of triangular structures in a network.   
Although there are some definitions of the clustering coefficient\cite{Watt1},\cite{Newm21},  
 we adopt the usual  global clustering coefficient $C_{(3)}$ \cite{Newm21} defined by 
%%%%%%%%%%%%%%%%%%%%%%%%%%%%%%%%%%%%%%%%%%%%%%%%%%%%%%%%%%%%%%%%%%%%%%%%%%%%%%%%%%%%%%%
\mathindent=37mm %%%%%%%%%%%%%%%%%%%%%%%%%%%%%%%%%%%%%%%%%%%%%%%%%%%%%%%%%%%%%
\begin{equation}
C_{(3)}=\frac{6\times \;number \;of \;triangles }{number \;of \;connected \;triplets }=\frac{ 6\Delta_3  }{\bar{S}_3},  
\end{equation}
where  $\Delta_q$ is generally the number of polygons with $q$ edges in a network. 
Some authors have made  extensions of the clustering coefficient for triangles to that for quadrilaterals.  
We, however, find it is difficult to extend it  further to that for circles with larger size.
But we  need to introduce  certain indices in order to uncover properties of general polygon structures in networks.  
From the expression of Eq.(1),  we can generalize it to $p$-$th $ clustering coefficient $C_{(p)}$ straightforwardly;  
\begin{equation}
C_{(p)}=\frac{2p\times \;number \;of \;polygons }{number \;of\; connected \;p\mbox{-}plets }=\frac{ 2p\Delta_p  }{\bar{S}_p}.
\end{equation}

\subsection{Adjacent Matrix Formulation}
We reformulate $C_{(p)}$ introduced in Eq.(2) by utilizing  an adjacent matrix $A=(a_{ij})$.     
Generally the powers, $A^2$,  $A^3$, $A^4$,  $\cdots$ of adjacent matrix $A$ give information as to respecting that a vertex connects other vertices  through $2,3,4, \cdots$ intermediation edges, respectively.   
The information of the connectivity between two vertices, $i_0$ and $i_n$, in $A^n$ also contains multiplicity of edges, generally.    
For resolving the degeneracy, we introduce a new series of matrices $R^n$ which give information as to respecting that a vertex connects other vertices  through $n$ intermediation edges without multiplicity. 
We can find it by the following formula\cite{Toyota3}; 
\begin{equation}
[ R^n] _{i_0i_n}=\displaystyle \sum_{i_1,\cdots,i_{n-1}} a_{i_0i_1} a_{i_1i_2}\cdots a_{{i-1},i_{n}} 
\frac{\displaystyle\prod_{i_k,i_j,i_k-i_j>1}^{n}(1-\delta_{i_ki_j})}{(1-\delta_{i_0i_n})}.
\end{equation}
   where the product of $(1-\delta_{i_ki_j})$ of the numerator plays role of protecting of degeneracies from strings and 
  $(1-\delta_{  i_0i_n})$ of the denominator is, however, needed to keep a closed string.

This expression has ($n-1$)-ply loops  in a computer program and so it is 
almost impossible to calculate $R^n$ within real time for large $N$.    
The expansion of Eq.(3) has $2^{n(n-1)/2}$ terms formally.  
 This value is $32768$ for $n=6$ that is needed for the analysis of six degrees of separation as will be discussed in the later section. 
Though many terms really vanish,  $R^6$ has still so complex expression.  
 We give the expressions of $R^1\sim R^6$;   
\mathindent=7mm
\begin{align}
[R^2]_{if} &=[A^2]_{if}-[A^2]_{ii}\delta_{if}=[A^2]_{if}-G_{if}, \nonumber \\
[R^3]_{if} &=[A^3]_{if}-\{ G,A \}_{if} +a_{if}, \nonumber \\
[R^4]_{if} &=[A^4]_{if}-\{ G,A^2 \}_{if} +\bigl\{A, diag(A^3)\bigr\}_{if} +2[A^2]_{if}  +[G^2-G-AGA]_{if}+3a_{if}[A^2]_{if} \nonumber \\
[R^5]_{if} &=[A^5]_{if} -\bigl\{A, diag(A^4)\bigr\}_{if}-\{ G,A^3 \}_{if} -\bigl\{A^2, diag(A^3)\bigr\}_{if}
+3\bigl([A^2]_{if}\bigr)^2 [A]_{if} \nonumber  \\
&+3[A^3]_{if}[A]_{if}+2\{ G^2,A \}_{if}+[GAG]_{if}- 6\{ G,A \}_{if} -\{ AGA,A \}_{if} +3[A^3]_{if}\nonumber \\ 
&   +\bigl\{A, diag(AGA)\bigr\}_{if}+2[diag(A^3G)]_{if} -[A\cdot diag(A^3)\cdot A]_{if} \nonumber - [diag(A^3)]_{if} \nonumber \\
&+3\sum_{k} a_{ik}a_{kf} \Bigl( [A^2]_{kf} + [A^2]_{ik}-\delta_{if}[A^2]_{kf} \Bigr) +4a_{if}\bigl( 1-a_{if} \bigr), 
%\{ G^2,A \}_{if} + [A^2]_{if}-[AGA]_{if}+[G^2-G]_{if}+3a_{if} [A^2]_{if}+a_{if}\bigl( [A^3]_{ii}+ A^3]_{ff}\bigr)\nonumber \\
%&=[A^4]_{if}-\{ G^2,A \}_{if}+[A^2]_{if}-[AGA]_{if}+[G^2-G]_{if}+3a_{if}[A^2]_{if}+\bigl\{A, diag(A^3)\bigr\}_{if}
\end{align}
where suffix is abbreviate in trivial cases and $\{\cdot, \cdot\}$ means the anti-commutation relation;  $\{A,B\}=AB+BA$.   
$diagA$   indicates the diagonal matrix  whose elements are the diagonal elements of $A$,  and 
$G$ is the diagonal matrix defined by
%%%%%%%%%%%%%%%%%%%%%%%%%%%%%%%%%%%%%%%%%%%%%%%%%%%%%%%%%%%%%%%%%%%%%%%%%%%%%%%%%%%%%%%
\mathindent=37mm %%%%%%%%%%%%%%%%%%%%%%%%%%%%%%%%%%%%%%%%%%%%%%%%%%%%%%%%%%%%%
\begin{eqnarray}
G&=&\left[
\begin{array}{cccc}
 k_1&0&0&\cdots \\
 0&k_2&0 &\cdots\\
 0&0&k_3 &\cdots\\
  \vdots &\vdots&\vdots&\ddots \\
   \end{array}
   \right] , 
\end{eqnarray}
where $k_i$ is the degree of vertex $i$. 

 $R^6$ is obtained after straightforward but long tedious calculations. 
We divide it into the following four parts to   brighten the prospects of the calculations.  
\mathindent=0mm

%\mathindent=7mm

\begin{align}
[R^6]_{if} =&\sum_{j,k,l,m,n} a_{ij} a_{jk}  a_{kl} a_{lm}  a_{mn} a_{nf} 
\Delta_{ik}  \Delta_{jl} \Delta_{km}  \Delta_{ln} \Delta_{mf}  \Delta_{il} \Delta_{jm}  \Delta_{kn}\Delta_{lf}  \Delta_{im} \Delta_{jn}  \Delta_{kf}\Delta_{in} \Delta_{jf} \nonumber \\
=&\sum_{j,k,l,m,n} a_{ij} a_{jk}  a_{kl} a_{lm}  a_{mn} a_{nf} 
 \Delta_{ik}  \Delta_{jl} \Delta_{km}  \Delta_{ln} \Delta_{mf}  \Delta_{il} \Delta_{jm}  \Delta_{kn}\Delta_{lf}  \Delta_{im} \Delta_{jn}  \Delta_{kf} \nonumber \\
&-\sum_{k,l,m,n} a_{if} a_{fk}  a_{kl} a_{lm}  a_{mn} a_{nf} \Delta_{ik}  \Delta_{fl} \Delta_{km}  \Delta_{ln} \Delta_{mf} \Delta_{il}  \Delta_{kn}  \Delta_{im} \nonumber \\
&-\sum_{j,k,l,m} a_{ij} a_{jk}  a_{kl} a_{lm}  a_{mi} a_{if}  \Delta_{ik}  \Delta_{jl} \Delta_{km}  \Delta_{li} \Delta_{mf}  \Delta_{jm} \Delta_{lf} \Delta_{kf}  \nonumber \\
&+\sum_{k,l,m} a_{if} a_{fk}  a_{kl} a_{lm}  a_{mi} \Delta_{ik}  \Delta_{fl} \Delta_{km}  \Delta_{li} \Delta_{mf},   \nonumber \\
\equiv & R^6[1]_{if} +  R^6[2]_{if}+R^6[3]_{if}+  R^6[4]_{if},
\end{align}
where $\Delta_{ik} = 1-\delta_{ik}$. 
Furthemore we divide $R^6[1]_{if}$ into the following four parts to   brighten the prospects of the caluculation. 

\begin{align}
 R^6[1]_{if} =&\sum_{j,k,l,m,n} a_{ij} a_{jk}  a_{kl} a_{lm}  a_{mn} a_{nf}  
\Delta_{ik}  \Delta_{jl}  \Delta_{km}   \Delta_{ln} \Delta_{mf}  \Delta_{il} \Delta_{jm}  \Delta_{kn}\Delta_{lf}  \Delta_{im} \Delta_{jn}  \Delta_{kf} \nonumber \\
 =&\sum_{j,k,l,m,n} a_{ij} a_{jk}  a_{kl} a_{lm}  a_{mn} a_{nf}   \Delta_{ik}  \Delta_{jl} \Delta_{km}  \Delta_{ln} \Delta_{mf}  \Delta_{il} \Delta_{jm}  \Delta_{kn} \Delta_{lf}   \Delta_{jn} \nonumber \\
-& \sum_{j,k,l,n} a_{ij} a_{jk}  a_{kl} a_{li}  a_{in} a_{nf}  \Delta_{ik}  \Delta_{jl} \Delta_{ln} \Delta_{if} \Delta_{kn}  \Delta_{lf} \Delta_{jn} \Delta_{kf}  \nonumber \\
-& \sum_{j,l,m,n} a_{ij} a_{jf}  a_{fl} a_{lm}  a_{mn} a_{nf}  \Delta_{if}  \Delta_{jl} \Delta_{fm}  \Delta_{ln} \Delta_{il}  \Delta_{jm}  \Delta_{jn} \nonumber \\
+& \sum_{j,l,n} a_{ij} a_{jf}  a_{fl} a_{li}  a_{in}  a_{nf} \Delta_{if}  \Delta_{jl} \Delta_{ln}  \Delta_{jn} \Delta_{mf},   \nonumber \\
 \equiv & R^6[1,1]_{if} +  R^6[1,2]_{if}+R^6[1,3]_{if}+  R^6[1,4]_{if}. 
\end{align}

The four terms are respectively expressed as follows;

\begin{align}
 R^6[1,1]_{if} =&[A^6]_{if}+[A^4]_{if} \bigl(4-(k_i+k_f) \bigr) 
+[AGA]_{if}(k_i+k_p)-\{AGA,A^2 \}_{if}-[A^2GA^2]_{if} 
 \nonumber \\
 &+2[A(G^2-3G)A]_{if}  +3\sum_{j,k} a_{ij}a_{jk}a_{kf}[A^2]_{jk}-\sum_j [A^3]_{jj} 
   \bigl(  a_{ij}[A^2]_{jp}+ [A^2]_{ij} a_{jf}\bigr) \nonumber \\
+&2\sum_j [A^2]_{ij} [A^2]_{jf} \bigl(  a_{ij}+ a_{jf}\bigr) 
+ [A^2]_{if} \bigl( k^2_i+k^2_f -3(k_i+k_f)+4\bigr) \nonumber \\
- &[A^3]_{if}\bigl(  [A^3]_{ii}+ [A^3]_{ff} \bigr) + \bigl([A^3]_{if} \bigr)^2 
  +  \sum_{j} a_{ij}a_{jf}\Bigl( \bigl( [A^3]_{ij} + [A^3]_{fj}\bigr)\nonumber \\
 -& [A^4]_{jj} -2\bigl( [A^2]_{ij} + [A^2]_{fj}\bigr)+ [AGA]_{jj}  
+ \bigl( ([A^2]_{ij})^2 + ([A^2]_{fj} )^2 \bigr) \Bigr) \nonumber \\
+&\Delta_{if}  \Biggl(   [A^2]_{if} \bigl(  (k_i-1)(k_f-1) +1-  [A^2]_{if} \bigr) 
-\bigl( [A^3]_{if} \bigr)^2  +\sum_j [A^2]_{ij} [A^2]_{jf} \bigl(  a_{ij}+ a_{jf}\bigr)    \nonumber \\
 +&  \sum_{j} a_{ij}a_{jf}\Bigl( \bigl( ([A^2]_{ij})^2 + ([A^2]_{fj} )^2 \bigr)  
-\bigl( [A^2]_{ij} + [A^2]_{fj}\bigr) \Bigr) \Biggr) \nonumber \\
+ & a_{if}  \Biggl(  [A^3]_{ff} \bigl(2k_f+k_i-5\bigr)  
+[A^3]_{ii}  \bigl(2k_i+k_f-5\bigr) +[A^2]_{if} \bigl(11-3k_i-3k_f\bigr)  \nonumber \\
& -2 \sum_{j} a_{ij}a_{jf}\Bigl( \bigl( [A^2]_{ij} + [A^2]_{fj}\bigr) \Bigr)  \Biggr), \nonumber \\
R^6[1,2]_{if} +&R^6[1,3]_{if} = -\Delta_{if} \Biggl(  [A^2]_{if} \bigl( [A^4]_{ii}+ [A^4]_{ff} \bigr) 
+4[AGA]_{if} -\{ A^2, G^2-3G \}_{if} -\{ AGA, A^2 \}_{if} \nonumber \\ 
-4&  [A^2]_{if}  -\sum_{j}  a_{ij} a_{jf}  \biggl( \Bigl([A^2]_{if})^2+  ([A^2]_{if})^2\Bigr) 
+2\bigl( [A^3]_{ij} + [A^3]_{fj}\bigr) -\bigl( [A^2]_{ij} + [A^2]_{fj}\bigr)          \biggr)  \Biggr)  \nonumber \\
+&a_{if} \Biggl( -2[A^2]_{if} [A^3]_{if} +2[A^2]_{if}(k_i+k_f-3) 
+2 \sum_{j} a_{ij}a_{jf} \bigl( [A^2]_{ij} + [A^2]_{fj}\bigr) \Biggr),
\nonumber \\
R^6[1,4]_{if} =&   [A^3]_{if}\Delta_{if} \Bigl(  ([A^3]_{if})^2 - 3  [A^2]_{if} +2    \Bigr). 
\end{align}

$R^6[2]_{if} $, $R^6[3]_{if} $ and $R^6[4]_{if}  $ are respectively given by the following expressions;
\begin{align}
R^6[2]_{if} +&R^6[3]_{if} = a_{if} \Biggl(  2[A^4]_{if} -( \bigl( [A^5]_{ii} + [A^5]_{ff}\bigr)  
-7\bigl( [A^3]_{ii} + [A^3]_{ff}\bigr)  +22[A^2]_{ij}  \nonumber \\
& +4[A^3]_{if}[A^2]_{if} +2\bigl( [A^3]_{ii}k_i + [A^3]_{ff}k_f\bigr)+\sum_{j} [A^3]_{jj} \bigl(a_{jf}+a_{ij}\bigr)  
\nonumber \\
 &-4\sum_{j} a_{ij}a_{jf} \bigl( [A^2]_{ij}    + [A^2]_{fj}\bigr)  -6\{A^2,G \}_{if} -2[AGA]_{if} +\{A,AGA\}_{ii}+\{A,AGA\}_{ff}  \Biggr),  \nonumber \\
R^6[4]_{if} =&  a_{if} \Biggl( [A^4]_{if} - [AGA]_{if} -\{A^2,G\} +5 [A^2]_{if}  - \Bigl( [A^3]_{ii} +  [A^3]_{ff} \Bigr) \Biggr). 
\end{align}

By unifying all the terms, we obtain the full expression of $R^6$.  
Lastly we give the expressions of Tr $R^n$ appearing in Eq. (7).  

 \begin{align}
Tr (R^2) &=0,   \nonumber  \\
Tr (R^3)  &=Tr (A^3),  \nonumber  \\
Tr(R^4)   &= Tr(A^4)-3 Tr(GA^2), +2Tr(A^2) + Tr(G^2-G), \nonumber \\
Tr(R^5)   &= Tr(A^5)-3 Tr(GA^3) +6 Tr(A^3) -diag(A^3) Tr(A^2)  +Ndiag(2A^3G-A^3),  \nonumber \\
Tr(R^6)&= Tr(A^6) +6Tr(A^4)-5Tr(GA^4) -4Tr(A^3)  +Tr(A^2G^2)  -6Tr(A^2G)+4Tr(A^2) \nonumber \\
&+2Tr(AGAG)   -\sum_i (a_{ii})^2 -\sum_{i,j} [A^3]_{jj}a_{ij}[A^2]_{ij} + 6  \sum_{i,j} a_{ij}[A^2]_{ij} +\sum_{i,j,k} a_{ij} a_{jk}  a_{ki}  [A^2]_{jk}.  
\end{align}
\mathindent=37mm
By using $R^n$, $\bar{S}_p$ and  generalized $p$-th clustering coefficient $C_{(p)}$ are given by
 \begin{equation}\bar{S}_p=\sum_{i,j} (R^{p-1})_{ij}/2,\end{equation}
\begin{equation} 
C_{(p)}=\frac{\mbox{Tr} R^p  }{ \displaystyle \sum_{i,j}R^{p-1}}, 
\end{equation}
where the denominator and the numerator indicates the contribution from open strings and a closed string, respectively.  
Thus usual clustering coefficient $C_{(3)}$ becomes 
\begin{equation}\displaystyle 
C_{(3)}=\frac{\mbox{Tr} R^3  }{ \displaystyle \sum_{i,j} (A^2)_{ij}-(A^2)_{ij} \delta_{ij}  }=\frac{ \mbox{Tr} A^3  }{ 
 ||A||-\mbox{Tr} A^2 }.
\end{equation}
where we introduced a new symbol $|| \cdots ||$ which denotes   $ ||A|| \equiv \sum_{i,j}A_{ij} $.

\section{Application to Six Degrees of Separation}
We analyze general $q$-$th$ degrees of separation based on the formalism developed  in the section 2.  
Aoyama has proposed a condition, so-called Milgram Condition, for $q$-$th$ degrees of separation\cite{Aoyama}; 
\begin{equation}
M_{q+1} \equiv \frac{\bar{S}_{q+1}}{N} \sim O(N).
\end{equation}

For six degrees of separation, we obtain from Eq.(6)
\begin{equation}
\bar{S}_7= \sum_{i,j} (R^6)_{ij}/2. 
\end{equation}

We investigate $q$-th degrees of separation by using Eq.(4)-(10) and the Milgram Condition.    
Here we place the focus on scale free networks where the degree distribution is 
$P(k)\sim k^{-\gamma}$. 
The networks can be constructed based on the configuration model \cite{Bebe}.\cite{Bend},\cite{Moll}    
which can systematically produce networks with arbitrary degree distribution. 
But the networks produced by the model are degenerate multigraphs, generally. 
We modify it a little to produce networks without multiple edges.  
Since it is not essential in this article, we omit the technical details of it. 
 Although  Eq. (3) reduces to Eq.(4)-(9),  we can not estimate the Milgram condition in large scale networks because of considerable computational complexity.  
 We can see that the results are stable and reliable while estimations are carried out  in small networks, 
   
Fig. 2 shows the relation between $\log_{10} M_q/N$ and $q$ for some $\gamma$'s  where the average 
degree $\langle k \rangle$ is four and network size $N=200$.   
$M_q/N$ increases linearly for every $\gamma$ with $q$. 
The interior of a rectangle in Fig.2 shows the region where the Milgram condition is satisfied.   
From Fig.2, while we see the four degrees of separation in  networks with $\gamma \leq 2.5$,  
we cannot recognize that vertecis are linked together in networks  with $\gamma \geq 3.5$ 
up to six degree of separation.  
$\gamma=2.75$ shows five degrees of separation and $\gamma=3.0$, in which many real-world networks have this value of exponent,  just shows six degrees of separation.  

 Comparing these results  with Aoyama's ones \cite{Aoyama} where we represent the median  of the region,   
there is a little difference between both results as shown in Table.1. 
Especially, it seems like Aoyam's assertion that $\gamma=2$ is a critical point for two degrees of separation  
conflicts with our result.  
But Aoyama  gives only a region 
where a separation number exists for every $\gamma$ and so we take the medians  of the region in Table 1.    
By considering moreover that Aoyama's calculations  are based on a tree approximation and thus the separation number $q$ is only a estimated one, two results are not necessarily inconsistent.   
Furthermore the estimations depend on how we build up networks,  in spite of  networks with the same $\gamma$. 

The fact our result comes closer to Aoyama's one\cite{Aoyama} for smaller $N$ (we do not go into the details),  is consistent with Aoyam's assertion\cite{Aoyama} that the accuracy of his calculations are decreased for larger $\gamma$. 

%%%%%%%%%%%%%%%%%%%%%%%  
\begin{figure}[t]
\begin{center}
\includegraphics[scale=1.2,width=12cm,height=9cm,clip]{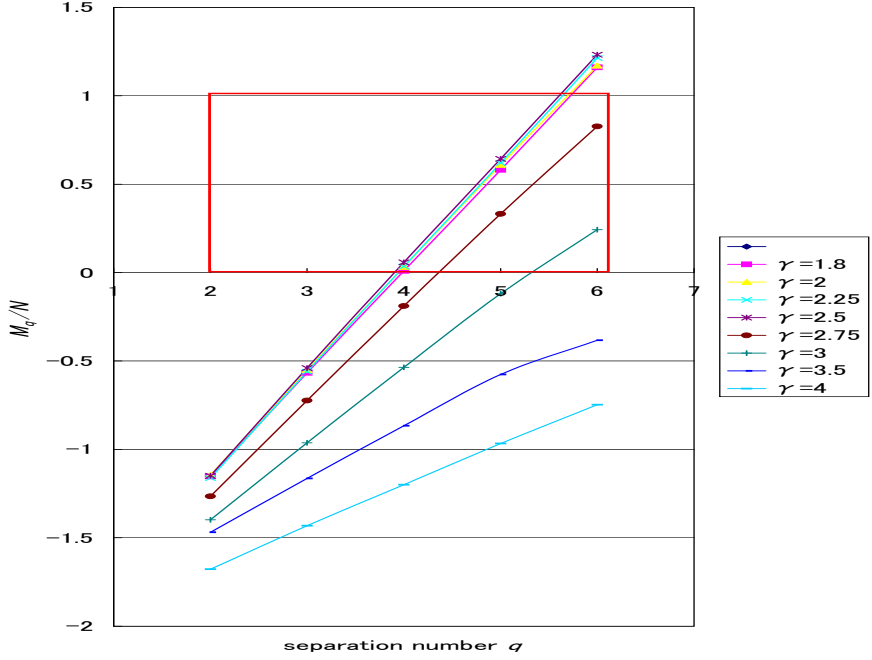}
\caption{ Separation number $q$ v.s. $M_q$ for scale free networks with  several $\gamma$}%%%%%%%%%%%%%%%%%%%%
\end{center}
\end{figure} 

%\vspace{2cm}

%%%%%%%%%%%%%%%%%%%%%%% scale=0.4,width=6cm,height=2cm,clip
%\begin{figure}[h]
%\begin{center}
%\includegraphics[width=6cm,height=2cm,clip]{Table1.eps}
%\caption{ Separation number $q$ v.s. $M$ for SFnet }%%%%%%%%%%%%%%%%%%%%
%\end{center}
%\end{figure} 
\begin{table}
\begin{center}
\begin{tabular}{|c|c|c|c|c|c|}\hline
$\gamma$ &2&2.5&2.75&3&3.5 \\\hline\hline
 Our results &4&4&5&6&× \\\hline
Aoyama's results &2&3&4&4&× \\\hline
\end{tabular}
\caption{ Comparison of our and Aoyama's  $q$  for diverse $\gamma$ }
\end{center}
\end{table}

We can demonstrate the validity of our results by directly evaluating the ratio $r$ of together connected vertices to  whole vertices from the power of an adjacent matrix, since  the network size $N$   is small.  
Fig.3 shows the relation between $q$ and $r$ for every $\gamma$. 
When every node connects with $50\%\sim60\%$ of vertices in a network,  it may be claimed in general that the network is  almost connected. 
Taking $r>50 \%\sim 60\%$ as  a borderline,  $q$ values derived from it are consistent with those estimated from $M_q$ in our calculation.   
Thus the point where $M_q/N$ becomes $O(1)$ really shows that a majority  of the vertices in a network  
connect each other.  

%%%%%%%%%%%%%%%%%%%%%%% 
\begin{figure}[t]
\begin{center}
\includegraphics[scale=1.0,width=9cm,height=6cm,clip]{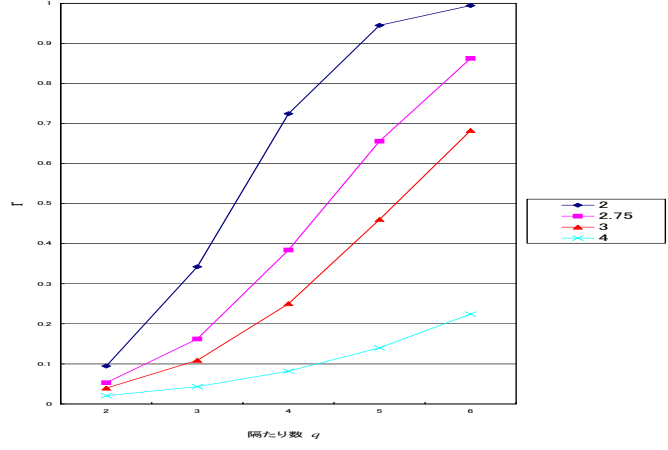}
\caption{ Separation number $q$ v.s. $r$ for scale free networks }%%%%%%%%%%%%%%%%%%%%
\end{center}
\end{figure} 

 \begin{figure}[t]
\begin{center}
\includegraphics[scale=1.0,width=8cm,height=6cm,clip]{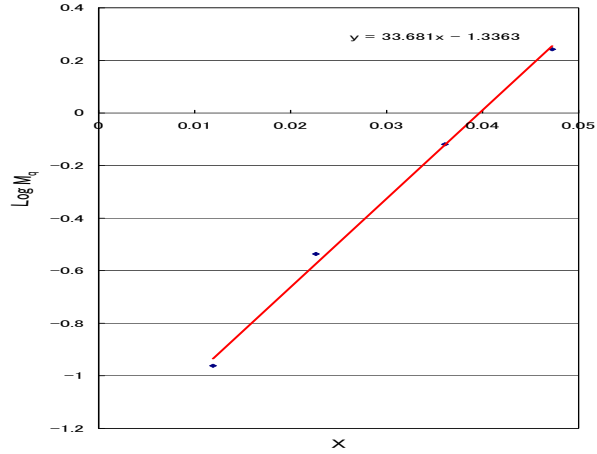}
\caption{ The sum of $C_{(p)}$ and $M_q$ for the scale network with $\gamma=3.0$ }%%%%%%%%%%%%%%%%%%%%
\end{center}
\end{figure} 

\section{Milgram Condition and Generalized Clustering coefficient}
We explore the relation between Milgram condition and the generalized clustering coefficients in this section. 
By making it, we can analyze how circle structures in a network is related with a separation number $q$. 
We define the following two quantities;
\begin{eqnarray}
X &\equiv& \sum_{p=3}^{q} C_{(p)},\\ \nonumber 
Y &\equiv& \log_{10} M_q.
\end{eqnarray}
Fig.4 shows the relation between $X$ and $Y$ at $\gamma=3.0$ in the scale free network with  $N=200$. 
We can recognize that $Y$  increases linearly with $X$; 
\begin{equation}
Y =A X +B.
\end{equation}
Such a relation holds for $1.8 \leq  \gamma <4.0$ in common.   
That is to say, it becomes clear that there is the relation of an exponential function between  $M_q/N$ and the sum of generalized clustering coefficients;   
\begin{equation}
M_q \sim \exp ( c \sum_{p=3}^{q} C_{(p)}),
\end{equation}
where $c$ is a constant determined by $A$ and $B$. 
Thus the separation number $q$ depends greatly  on the sum of $C_{(p)}$( $p \leq q$ ), which represents 
the state of the circle structures up to $q$ in a network. 
This  indicates that the generalized clustering coefficient introduced in this article is an effective 
index to explore $q$-$th$ degrees of separation.

%%%%%%%%%%%%%%%%%%%%%%% 
 \begin{figure}[t]
\begin{center}
\includegraphics[scale=1.0,width=8cm,height=6cm,clip]{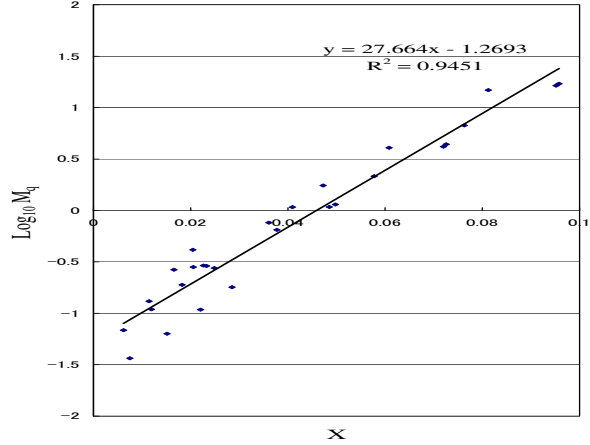}
\caption{ The sum of $C_{(p)}$ and $M_q$ for the scale free networks with $\gamma=2.0,2.25,2.5,2.75,3.0,3.5,4.0$ }%%%%%%%%%%%%%%%%%%%%
\end{center}
\end{figure}  
%%%%%%%%%%%%%%%%%%%%%%% 
%\begin{figure}[t]
%\begin{center}
%\includegraphics[scale=0.8,width=8cm,height=5cm,clip]{SF2Deg.eps}
%%\rotatebox{270}{ \includegraphics[scale=0.3,clip]{denomiNPo.eps} \hspace{15mm} }
%\hspace{5mm}
%\includegraphics[scale=0.8,width=8cm,height=5cm,clip]{SF2Deg2.eps}
%\end{center}
%\caption{$M_2$ ;upper side is plots when $\gamma \geq 2$ and underside is those of $\gamma=1$ and $N$ for scale free networks. } %%%%%%%%%%%%%%%%%%%%
 %\end{figure} 
 %%%%%%%%%%%%%%%%%%%%%%% 

We observe that further relations hold for  $q$ and $M_q/N$ by drawing a superposed diagram of the above-mentioned linear relation for diverse $ \gamma $'s. 
Fig.5 is the superposed diagram  for  $2.0 \leq  \gamma <4.0$. 
  The linear lines for  $2.0 \leq  \gamma <4.0$ almost are joined to be  a line with an almost common gradient.   
This means that $q$ depends only on the generalized clustering coefficient and not on $\gamma$, directly. 
Thus  the exponent in scale free networks is not crucial for the separation number but the state of circle structures in networks is essential. 

The reason why the relations holds is outstanding issue and only a phenomenological relation at present.

\section{Summary}
In this article, we first introduced the generalized clustering coefficient, which has information on the state of circle structures in a network, based on the string formulation proposed by \cite{Aoyama} to analyze networks.  
Fusing adjacent matrix $A$ into the formalism, we reformulate the string formalism to define the generalized $q$-$th$  clustering coefficient in a compact way\cite{Toyota3}, \cite{Toyota4}.     
Then we introduce the  $R$ matrix in the formalism developed in this article instead of $A$.  
The powers of $R$ play central role in the analysis of this article.    
The explicit representations of  $R^n$ for $n=2\sim6$ are given after straightforward but tedious calculations.   

Next we applied the formulation to the subject of $q$-$th$, especially $q=6$,  degrees of separation. 
We evaluated whether Milgram condition proposed by Aoyama's article holds or not for diverse exponents in scale free networks. 
We find that as the exponent $\gamma$ is larger, so it is more difficult that Milgram condition holds.    
 The six degrees of separation is just founded  at $\gamma=3$ whose value  is fairly universally observed in real-world networks.

We also find that the result seems to be  a little different from Aoyama's one\cite{Aoyama}. 
We think that it does not mean  necessarily inconsistency, considering that Aoyama's evaluation is based on tree approximation and furthermore the way to construct networks is maybe different (Aoyama does not explain the way to construct networks and the construction  of networks in this article include some original way in avoiding multiedges ).    Our results is also supported  by analyzing the number of zero-components in $A^n$.

The our construction is based on the configuration model\cite{Bebe}.\cite{Bend},\cite{Moll} with average degree $<k>=4$.  
According to some sociologists,  the estimated  average number of  acquaintances of a person is 290 \cite{Bernard1},\cite{Bernard2},\cite{Bernard3}. 
Considering this estimation, the separation number would really take  smaller values for every exponent.    

The following problems are yet left in future:
\begin{enumerate}
\item
Finding  explicit  expressions of $R^n$ for arbitrary $n$ by applying our formalism. 
 Then finding  a general formula for  $R^n$.  
  \item
Revealing relations between $q$-$th$ degrees of separation and $N$, $\langle k \rangle$ or $<k^n>$.    
 More definitely,  discovering the relations between $q$ and  $N$, $\langle k \rangle$ or $<k^n>$. 
   \item
The reason why the relations (18) holds is outstanding issue. 
So finding some theoretical reasons for  phenomenological relations between the  separation number and various circle structures, especially $C_{(q)}$.
 \item Attempt whether this relation holds or not in other networks, especially small world networks which can at least control the usual clustering coefficient by construction.   
 
\end{enumerate}

%\section*{Acknowledgments}

%\bibliographystyle{ieicetr}
%\bibliography{myrefs}

\begin{thebibliography}{99}

\bibitem{Newm21}M.E.J.Newman,"Ego-centered networks and the ripple effect or why all your friends are wired", Social Networks 25 (2003) p.83;arXiv. cond-mat/0111070

\bibitem{Aoyama} H. Aoyama, "Six degrees of separation; some caluculation", SGC library65, " Introduction to Network Science", (2008) in Japanese;
H, Aoyama, Y.Fujiwara, H, Ietomi, Y. Ikeda and W.Soma "EconoPhysics",Kyouritu Shuppan 2008 

\bibitem{Milg}S. Milgram, "The small world problem", Psychology Today 2, 60-67 (1967)
\bibitem{Milg2}J. Travers and S. Milgram, "An Experimental Study of the Small World Problem", Sociometry 32, 425 (1969) 
\bibitem{Milg3} C. Korte and S. Milgram, "Acquaintance edges between White and Negro populations: Application of 


\bibitem{Pool} I.S. Pool and M. Kochen,  "Contacts and Influence", Social Networks, 1(1978/1979)5-51(This paper was actually written in 1958)

\bibitem{Watt1}D. J. Watts　and S. H. Strogatz, "Collective dynamics of 'small-world' networks",　Nature,393, 440-442(1998)
\bibitem{Watt2}D. J. Watts, "Six degree-- The science of a connected age", W.W. Norton and Company, New York (2003)

\bibitem{Albe2}A.-L.Barabasi and R.Albert, "Emergence of scaling in random networks", Science, 286, 509-512(1999)

\bibitem{Albe3}A.-L.Barabasi and R.Albert, "edgeed: The New Science of Networks",  Perseus Books Group (2002) 
edgeed: How Everything Is Connected to Everything Else and What It Means for Business, Science, and Everyday Life
Plume ; ISBN: 0452284392 ; Reissue 版 (2003/04/29)
\bibitem{Albe1}R.Albert and A-.L. Barabasi, "Statistical Mechanics of complex networks",Rev. Mod. Phys. 74, 47-97(2002)

\bibitem{Klein} J.S. Kleinfield, "The small world problem", Society 39(2) pp.61-66(2002):
"COULD IT BE A BIG WORLD? ", http://www.uaf.edu/northern/big$ \_$world.html 

\bibitem{Newm}M.E.J. Newman, A.-L.Barabasi  and D. J. Watts, "The Structure and Dynamics of Networks", Princeton Univ. Press, 2006 
　
 \bibitem{Doro1}S. N. Dorogovtsev, A.V. Goltsev and J.F.F. Mendes, "Pseudo fractal scale-free web", Phys. Rev. E.65, 066122(2002) 
\bibitem{Doro2}S. N. Dorogovtsev and J.F.F. Mendes, "Evolution of Networka", Oxford Univ. Press, Oxford(2003)

\bibitem{Watt4}D. J. Watts et al., Small World Project-Columbia University. 
http://small world.columbia.edu/

\bibitem{Watt3}P.S.Dodds, R.Muhamad and D.J. Watts, "An Experimental Study of Research in Global \\Social Networks", Science 301, pp.827-829:  \\http://small world.columbia.edu/images/dodds2003pa.pdf (2003)

 \bibitem{Toyota1} N. Toyota, "Some Considerations on Six Degrees of Separation from A Theoretical Point of View”, arXiv:0803.2399
 \bibitem{Toyota2} N. Toyota, "Comments on Six Degrees of Separation based on the le Pool
and Kochen Modelsgendary", arXiv:0905.4804

%\bibitem{Toyota4}N. Toyota and T. Sakamoto, to be appeared. 


\bibitem{Toyota3} N. Toyota, IEICE Thecnical Report, "String Formalism for $p$-Clustering Coefficient-Toward Six Degrees of Separations",NLP2009-49(2009) in Japanese. 
\bibitem{Toyota4} N. Toyota, " $p$-th Clustering coefficients $C_{p}$ and Adjacent Matrix for Networks: Formulation based on String", arXiv:0912.2807

\bibitem{Bebe}A.Bebessy, P.Bebessy and J. Komlos, Stud,. Sci., Math. Hangary, 7343- 7353 (1972)
\bibitem{Bend}E.A.Bender and E.R. Candield, J. Comb. Theory A. 24. 296-307 (1978)
\bibitem{Moll} M. Molloy and B. Reed, Comb., Prob. and Compt. 6. 161-179 (1995);  7. 295-305 (1998


\bibitem{Erdos1}P. Erdos and A. Renyi," On random graphs I", Publicationes Mathematicae Debrecen6, 290-297, 1959

\bibitem{Bernard1}P.D.Killwoth,E.C.Johnsen, H.R.Bernard, G.A.Shelley and "Estimating the size of personal networks", Social Networks 12,289-312 (1990)
\bibitem{Bernard2} H.R.Bernard, E.C.Johnsen, P.D.Killwoth and S. Robinson, " Estimating the size of average  personal network and of an event population; Some empirical results", Social Science Research 20, 109-1211991)
\bibitem{Bernard3} H.R.Bernard, P.D.Killwoth, E.C.Johnsen,  and  C.McCarty, " Estimating the ripple effect of a disaster", Connections 24(2), pp.16-22(2001)

\bibitem{Lind}P.G.Lind, M.C.Gonzalez and H.J.Hermann, "Cycles and clustering in bipartite networks", Phys.Rev.E 72,056127 (2005)
\bibitem{Zhang} P.Zhang, J.Wang, X.Li, M.Li, Z.Di and Y.Fan,"Clustering coefficient and community structure of bipartite networks", Physica A, 387, 6869-6875(2008) 






%\bibitem{Masu1}N. Masuda and N. Konno, " What is conplex networks", (Kodanshya 2006)
%\bibitem{Yuta1}湯田聴夫，小野田直亮、藤原義久、「ソーシャルネットワーキング・サービスのリンク特性とクラスター構造」、ネットワーク生態学研究会シンポジウム（2005）

%\bibitem{Newm2}M.E.J. Newman, "The structure of scientific collaboration networks", Proc. Natl. Acad. Sci. USA 98(2001)404;  %
%"A sutudy of scientific coauthorship networks, scientific collaboration networks I, Network consideration and fundamental results",
% Phys.Rev. E64, (2001)016131;
% "scientific collaboration networks II,Shortest path, weighted networks, and centrality",  Phys.Rev. E64, (2001)016132

%\bibitem{Albe4}A.-L.Barabasi, H. Jeong, Z.Neda, E.Ravasz, A Shubert and T. Vicsek, 
%" Evolution of social network of scientific collaborations", Physica A311, (2002)590

%\bibitem{Ferr} C. Ferrer I and R.V.Sole, "The small world of human language", Proc. R. Soc. London B268(2001)2261
%\bibitem{Fell} P.A. Fell and A. Wagner, "The small world of metabolism", Nature Biotechnology18(2000)1121

%\bibitem{Bacon} Patrick Reynolds and the CS Web Team,"  The Oracle of Bacon at Virginia",
 %  http://oracleofbacon.org/

%\bibitem{Barr1}A. Barrat and M. Weigt, "On the properties of small-world networks", Eur. Phys. J. B13, (2000)547

%\bibitem{Newm2}M.E.J. Newman, C. Moor and D. J. Watts, "Mean-field solution of small-world network model", 
%Phys. Rev. Lett. 84(14)3201



%


%\bibitem{Bian}G. Bianconi and  A.-L.Barabasi, "Competion and Multiscaling in evolving Networks", Europianphys. Lett., 54(4), 436(2001) 

%\bibitem{Klem1}K. Klemm, V.M. Eguiluz, "Highly clusterd scale-free networks", Phys. Rev. E.65, 036123(2002) 

%\bibitem{Klem2}K. Klemm, V.M. Eguiluz, "Growing scale-free networks with small world behavior", Phys. Rev. E.65, 057102(2002) 

 %\bibitem{Rava}E. Ravasz and  A.-L.Barabasi, "Hierarchial Organization in Complex Networks", Phys. Rev. E.67, 026112(2003) 

\end{thebibliography}

\end{document}